# Optimally Chiral Electromagnetic Fields: Helicity Density and Interaction of Structured Light with Nanoscale Matter

Mina Hanifeh[1], Mohammad Albooyeh, and Filippo Capolino[2]

Department of Electrical Engineering and Computer Science, University of California, Irvine, California, USA 92697

## ABSTRACT

We propose the concept of helicity maximization applicable to structured light and obtain a universal relation for the maximum of helicity density at a given field energy density. We further demonstrate that using structured light with maximized helicity density eliminates the need of the specific knowledge of energy and helicity densities in determining the chirality of a nanoparticle. The helicity maximization concept generalizes the use of the dissymmetry factor $g$ in chirality detection to any chiral structure light illuminating nanoparticles.

Chirality is observed in a plethora of phenomena ranging from molecular structures to spiral galaxies[1]–[4]. Its ubiquity is not the only reason for its allure since it is related to fundamental aspects of life sciences, including biochemistry, genetics, molecular biology, pharmacology, astronomy, etc. [5]–[11]. Therefore, probing chirality is of supreme importance. In this paper chirality at nanoscale is probed with light-matter interaction[8], [10]. Similar to electric permittivity $\epsilon$ and magnetic permeability $\mu$, a medium chirality is represented by $\kappa$, and the material constitutive relations $\mathbf{D} = \epsilon\mathbf{E} + i\kappa\sqrt{\mu_0\epsilon_0}\mathbf{H}$ and $\mathbf{B} = \mu\mathbf{H} - i\kappa\sqrt{\mu_0\epsilon_0}\mathbf{E}$ define the link between the electric and magnetic fields $\mathbf{E}$ and $\mathbf{H}$, and the electric displacement and magnetic induction vectors $\mathbf{D}$ and $\mathbf{B}$ [12], [13]. Here $\epsilon_0$ and $\mu_0$ are the vacuum permittivity and permeability, respectively, and we assume fields with time dependence $\exp(-i\omega t)$, where $\omega$ is the angular frequency.

Usually, chirality of a bulk material $\kappa$ is detected and characterized through circular dichroism (CD) experiments[14]–[16]. For example, exposing a test slab with thickness $d$ to two distinct circularly polarized (CP) plane waves with polarizations of opposite handedness, chirality is characterized via

$$\Im\{\kappa\} = \frac{1}{2k_0 d}\tanh^{-1}\left(\frac{|t^+|^2 - |t^-|^2}{|t^-|^2 + |t^-|^2}\right), \quad (1)$$

where $|t^\pm|$ are the measured transmission coefficients in the two experiments and $k_0$ is the wavenumber in vacuum. Chirality of a homogeneous bulk material may be the result of chirality of its individual chiral inclusions[15]. The response of chiral nanoparticles (NPs) (e.g. the inclusions of a bulk mixture, like molecules or engineered nanostructures) to an external electromagnetic field is modeled by induced electric and magnetic dipole moments $\mathbf{p}$ and $\mathbf{m}$, related to *local* electric and magnetic fields through the linear relations $\mathbf{p} = \alpha_\mathrm{ee}\mathbf{E} + \alpha_\mathrm{em}\mathbf{H}$ and $\mathbf{m} = \alpha_\mathrm{me}\mathbf{E} + \alpha_\mathrm{mm}\mathbf{H}$ (see Ref.[13]), where $\alpha_\mathrm{me} = -\mu_0^{-1}\alpha_\mathrm{em}$ due to reciprocity. Here, we assume bi-isotropic reciprocal NPs with individual electric, magnetic, and magnetoelectric polarizabilities $\alpha_\mathrm{ee}$, $\alpha_\mathrm{mm}$, and $\alpha_\mathrm{em}$, respectively, where $\alpha_\mathrm{em}$ indicates the NP's chirality. These individual NP's polarizabilities are related to the bulk optical properties of the mixture $\epsilon$, $\mu$, and $\kappa$ [17]. For example, for a dilute mixture these relations are $\epsilon = \epsilon_0 + N\alpha_\mathrm{ee}$, $\mu = \mu_0 + N\alpha_\mathrm{mm}$, and $\kappa = -iN\alpha_\mathrm{em}/\sqrt{\mu_0\epsilon_0}$, where $N$ is the number of NPs in the unit volume. Therefore, the magnetoelectric polarizability of an individual NP can be *indirectly* obtained using Eq. (1) for a dilute mixture after the characterization of the bulk chirality parameter $\kappa$ [17]. However, determining the magnetoelectric polarizability $\alpha_\mathrm{em}$ of individual chiral NPs in a host medium is challenging. For example, the absorption bands of the host and chiral inclusions may overlap, or for a dense mixture the above simple relations are not valid[17], [18]. Moreover, limited dynamic ranges of measuring devices, small amount of mixture solution, or weak chirality of individual NPs [19] impose challenges in chirality detection of small NPs in conventional CD experiments.

---

[1] Mhanifeh@uci.edu

[2] F.capolino@uci.edu

Several attempts have been made to improve chirality detection by exploiting chiral fields, e.g., standing waves, near-fields, etc. [19]–[36]. However, the maximum capability of such fields in chirality detection has never been explored. Besides, the determination of the strength of material chirality independent of field properties, i.e., *characterization*, is not investigated. Subsequently, we found it essential to address these issues by a rigorous analysis of light-matter interaction for optically small NPs and to find the kinds of fields and the required conditions that optimize chirality detection and also enable characterization.

Here we employ two different excitations of a NP to detect its chirality through probing extinction powers, e.g., using optical theorem[37]. As discussed in [38], *differential extinction* is the difference of the extinction power related to these two excitations,

$$\Delta P_{ext} = P_{ext}^+ - P_{ext}^-, \quad (2)$$

and is an appropriate measurable choice for NPs' chirality detection. Each term is defined as $P_{ext}^\pm = \omega \Im\{\mathbf{p}^\pm \cdot (\mathbf{E}^\pm)^* + \mu_0 \mathbf{m}^\pm \cdot (\mathbf{H}^\pm)^*\}/2$ where "*" denotes complex conjugation. The choice of the "+" and "–" signs for discriminating the two experiments is dictated by the signs of helicity densities $h^\pm = \Im\{\mathbf{E}^\pm \cdot (\mathbf{H}^\pm)^*\}/(2\omega c_0)$ of the fields used in each of them (see FIG 1).

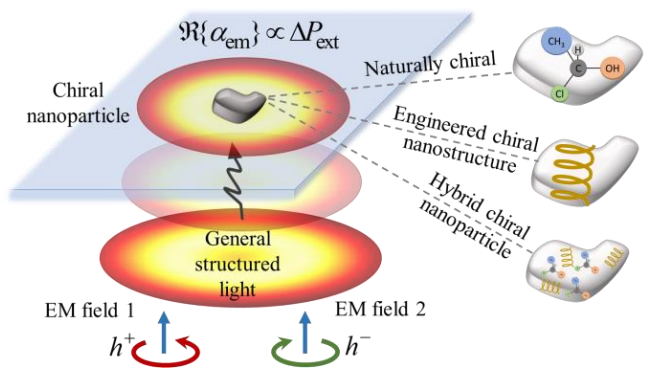

FIG 1. Illustration of the method to reveal the NP's chirality using two general electromagnetic fields (EM fields 1 and 2), with structured light with opposite handedness (denoted by the helicity densities $h^\pm$) at the position of the NP. The chiral particle may be natural, engineered, or a hybrid of these two.

Now, by defining $\Delta u_e = u_e^+ - u_e^-$, $\Delta u_m = u_m^+ - u_m^-$, and $\Delta h = h^+ - h^-$, respectively, as the differences between the time-averaged electric energy $u_e^\pm = \epsilon_0 |\mathbf{E}^\pm|^2/4$, magnetic energy $u_m^\pm = \mu_0 |\mathbf{H}^\pm|^2/4$, and helicity $h^\pm = \Im\{\mathbf{E}^\pm \cdot (\mathbf{H}^\pm)^*\}/(2\omega c_0)$ densities for fields in the two experiments, the differential extinction $\Delta P_{ext}$ reads

$$\Delta P_{ext} = 2\epsilon_0^{-1}\Im\{\alpha_{ee}\}\Delta u_e + 2\Im\{\alpha_{mm}\}\Delta u_m + \omega c_0 \Re\{\alpha_{em}\}\Delta h. \quad (3)$$

Therefore, chirality of a NP is detectable from the differential extinction $\Delta P_{ext}$ if the energy densities of the fields in the two experiments are identical, i.e., if they fulfill the condition

$$\Delta u_e = \Delta u_m = 0, \quad (4)$$

which means that the structured lights are such that $|\mathbf{H}^+|=|\mathbf{H}^-|$ and $|\mathbf{E}^+|=|\mathbf{E}^-|$ at the NP location. Applying condition (4) into Eq. (3) leads to a concise relation between the differential extinction and the helicity density of the fields and the chirality of the NP, i.e.,

$$\Delta P_{ext} = \omega c_0 \Delta h \Re\{\alpha_{em}\}. \quad (5)$$

For a reliable detection of chirality of a NP, the differential extinction $\Delta P_{ext}$ should be higher than the noise level of a measuring device. It is clear from Eq. (5) that for a given minimum value of measurable differential extinction, higher differential helicity density $\Delta h$ corresponds to the detection of NPs with weaker chirality strength. Therefore, as a major goal to maximize detection possibility of chirality, assuming the field choice satisfies condition (4), one requires to maximize the differential helicity density $\Delta h$, which implies

$$h^+ = -h^- = h. \quad (6)$$

Moreover, we need to determine what fields offer the maximum value for helicity density $h$, which we refer them as "best" scenarios for chirality detection. To that end, let us denote the electric and magnetic fields by $\mathbf{E}=|\mathbf{E}|\hat{\mathbf{e}}$ and $\mathbf{H}=|\mathbf{H}|\hat{\mathbf{h}}$ which results in $h=|\mathbf{E}||\mathbf{H}|\Im\{\hat{\mathbf{e}}\cdot\hat{\mathbf{h}}^*\}/(2\omega c_0)$, where $\hat{\mathbf{e}}$ and $\hat{\mathbf{h}}$ are complex dimensionless unit vectors. This means that the two components $\Im\{\hat{\mathbf{e}}\cdot\hat{\mathbf{h}}^*\}$ and $|\mathbf{E}||\mathbf{H}|$ which are linked to the polarization and the amplitude of the field, must be simultaneously maximized in order to maximize the helicity density $h$. While the maximum value of $|\Im\{\hat{\mathbf{e}}\cdot\hat{\mathbf{h}}^*\}|$ is unity, that of the term $|\mathbf{E}||\mathbf{H}|$ requires more elaborations since the field amplitudes can be unconditionally increased. Therefore, we consider all structured light such that the total energy density $u=(\epsilon_0|\mathbf{E}|^2+\mu_0|\mathbf{H}|^2)/4$ is kept constant at the NP location. Employing the method of Lagrange multipliers [39] the maximum of $|\mathbf{E}||\mathbf{H}|$, under the constraint of constant energy density $u$, occurs when $|\mathbf{E}|=\eta_0|\mathbf{H}|$.

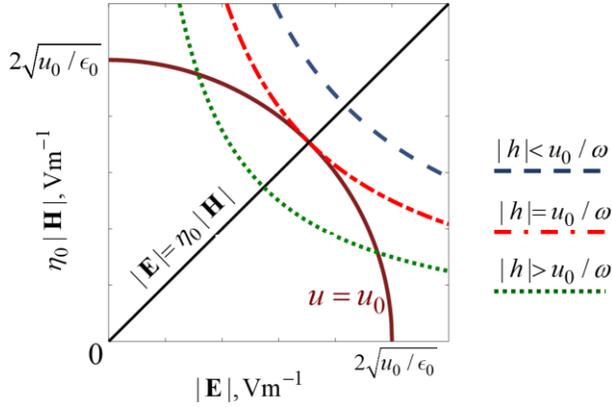

FIG 2. The normalized magnetic to electric field amplitude curves for a general electromagnetic field. The red solid curve demonstrates fields with constant energy densities while the blue dashed lines show fields with constant helicity densities. The black solid line represents fields with $|h|=u/\omega$ that also shows maximum helicity density with a given energy density.

To visualize this important result, the circle $4\epsilon_0^{-1}u_0 = |\mathbf{E}|^2 + |\eta_0\mathbf{H}|^2$, for a typical value of the energy density $u_0$, is plotted in the plane of $|\mathbf{E}|$ and $|\eta_0\mathbf{H}|$ in Fig. 2 (red solid line). Moreover, assuming $\Im\{\hat{\mathbf{e}}\cdot\hat{\mathbf{h}}^*\} = \pm 1$ is satisfied, three hyperbolic curves, which are the loci of $|\mathbf{E}|$ and $|\eta_0\mathbf{H}|$ satisfying $|h|=|\mathbf{E}||\mathbf{H}|/(2\omega c_0)$ for three magnitude values of the helicity density $|h|$, are plotted in the same figure. The red circle is tangent to the hyperbola with $|h_0|=u_0/\omega$, and the tangent point is where $|\mathbf{E}|=\eta_0|\mathbf{H}|$ at the hyperbola's vertex. Therefore, we claim that any structured light that locally satisfies the two conditions

$$\begin{cases} \Im\{\mathbf{E}\cdot\mathbf{H}^*\} = \pm|\mathbf{E}||\mathbf{H}| \\ |\mathbf{E}| = \eta_0|\mathbf{H}| \end{cases} \quad (7)$$

possesses the maximum available helicity density attributed to its given energy density. We call fields satisfying conditions (7) *optimally chiral*. It is noteworthy that in quantum electrodynamics, helicity defined as the projection of the angular momentum onto the linear momentum, is considered as an operator with eigenvalues $\pm 1$ [40]. Moreover, it is interesting that optimally chiral are eigenstates of the helicity operator referred as the Riemann-Silberstein linear combinations $\mathbf{G}_\pm = (\mathbf{E}\pm i\eta_0\mathbf{H})/\sqrt{2}$ [41], [42].

Conditions (7) is obtained for general electromagnetic fields. A plane wave with circular polarization is a trivial example of optically chiral field. However, since the energy and helicity densities of plane waves are not localized they are not appropriate candidates for chirality detection when dealing with NPs with weak chirality[20]. But the generality of (7) paves the way to conceive many other optimally chiral fields for chirality detection. Therefore, with the goal of enhancing the possibility of chirality detection of NPs, we propose optimally chiral structured light configurations, with localized energy and helicity. For example, the commonly used circularly polarized Gaussian beams are optimally chiral under paraxial approximation near the beam axis (see Ref. [38] for detailed discussions). To investigate a nontrivial "best" scenario, we consider an optical vortex beam with unique properties which provide intriguing applications in microscopy and spectroscopy[43].

In order to acquire an optically chiralbeam, we propose the superposition of two vortex beams propagating along the $z$ direction, i.e., an azimuthally polarized beam (APB) with electric field $\mathbf{E} = V_A e^{i\psi} f_\varphi(\rho,z)\hat{\boldsymbol{\varphi}}$ and a radially polarized beam (RPB) with magnetic field $\mathbf{H} = (V_R/\eta_0)f_\varphi(\rho,z)\hat{\boldsymbol{\varphi}}$, where $f_\varphi$ is the field profile as a function of radial $\rho$ and longitudinal $z$ positions, and $\hat{\boldsymbol{\varphi}}$ is the unit vector in the azimuth direction $\varphi$. Each beam is obtained using the superposition of two Laguerre-Gaussian beams[44], [45]. Moreover, $V_A$ and $V_R$ are, respectively, the APB and RPB field amplitudes and $\psi$ is their phase difference. The electric and magnetic field components (including the longitudinal ones) in the introduced combination, which we call ARPB hereafter, are always parallel, with a controllable phase shift $\psi$. In the following we investigate the helicity density on the beam axis, where only the $H_z = -V_A\eta_0^{-1}e^{i\psi}f_z(\rho,z)$ and $E_z = V_R f_z(\rho,z)$ field components are present, and $f_z = (ic_0\omega^{-1}\rho^{-1})\partial(\rho f_\varphi)/\partial\rho$ [44]–[46]. According to conditions (7), helicity density $h = -V_A V_R \eta_0^{-1}|f_z|^2_{\rho,z=0}\sin(\psi)/(2\omega c_0)$ is maximized if $\psi = \pi/2$ and $V_A = V_R$. To verify, we depict in Fig. 3 the normalized (to the maximum) helicity density when the ratio $V_A/V_R$ and phase shift $\psi$ are varying so that energy density $u_0 = \epsilon_0|f_z|^2_{\rho,z=0}(|V_A|^2 + |V_R|^2)/4$ remains constant.

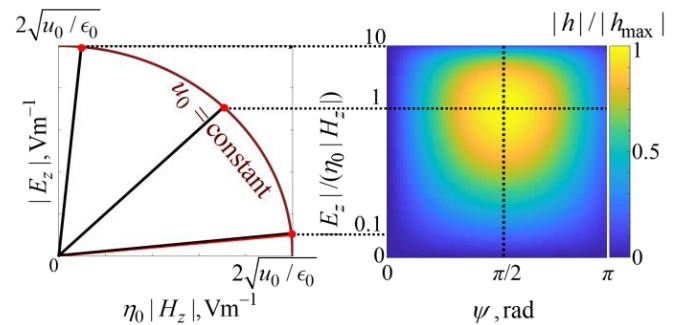

FIG 3. Normalized (to maximum) helicity density of an ARPB, i.e., the superposition of two structured beams, an APB and an RPB, evaluated for different relative amplitudes of the two beams and their relative phase shift $\psi$. Helicity is evaluated on the beam axis

at the minimum beam-waist plane (i.e., for $\rho = 0, z = 0$) and all results are based on assuming constant total energy density $u$ at the same location. The maximum of the helicity density occurs when two conditions $|E_z| = \eta_0 |H_z|$ and phase shift $\psi = \pi/2$ are met.

It is shown in [38] that energy and helicity densities of optimally chiral beams are enhanced when they are focused in smaller spots, however, such an enhancement is limited due to diffraction. Therefore, next we show that near-fields of nanoantennas (NAs) can be optically chiral at subwavelength scale, which would lead to an improvement in chirality detection of small NPs. As we proved in[38], the scattered near and far fields of an optically small nanoantenna, which is modeled by electric $\mathbf{p}^{NA}$ and magnetic $\mathbf{m}^{NA}$ dipole moments, under the "balanced" condition

$$\mathbf{p}^{NA} = \pm i\mathbf{m}^{NA}/c_0, \tag{8}$$

are optimally chiral. Here we demonstrate that appropriately engineered dielectric NAs with high refractive index, which provides electric and magnetic Mie resonances[38], with equivalent dipoles that satisfy (8), are convenient candidates for helicity-enhancement. For a NA, conditions (8) and hence (7) are satisfied when the electric polarizability $\alpha_{ee}^{NA}$ and magnetic polarizability $\alpha_{mm}^{NA}$ are "balanced", i.e., when $\alpha_{ee}^{NA} = \epsilon_0 \alpha_{mm}^{NA}$, and the illuminating external field satisfies conditions (7).like CP beams, the ARPB, etc.

A NA with "balanced" condition (8) can be made of a Si nanosphere located on the axis of an ARPB traveling along the $+z$ direction. The normalized (to the incident) helicity density of the total nearfield of such a NA is depicted in FIG 4, normalized to the helicity of the incident ARPB field.

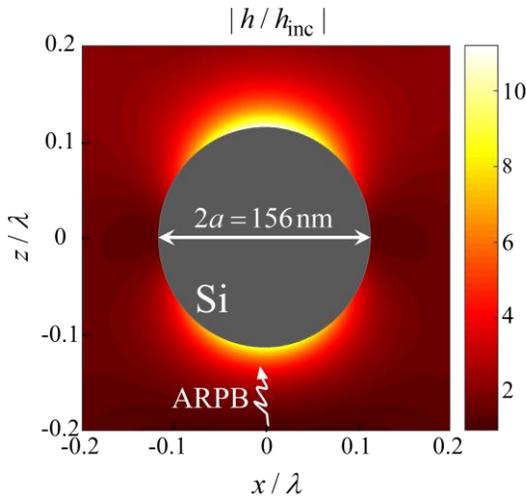

FIG 4. Helicity density enhancement $|h/h_{inc}|$ around a Si nanoantenna illuminated by an optimally chiral ARPB with $V_A = V_R = 1V$ and $\psi = \pi/2$ with free space wavelength $\lambda = 680$ nm, and beam parameter $w_0 = \lambda$ (see more details in [44], [45] and [38]). The sphere with radius $a = 78$ nm is assumed to be made of silicon with refractive index of $n = 3.807 - i0.014$ at the operational wavelength[47].

We observe more than an order of magnitude enhancement in the local helicity density in proximity of the sphere, along the beam axis which drops to values around 2-3 at 50 nm away from the surface. This near field with enhanced helicity density has the remarkable potential to improve detection of chiral NPs. Note that the retrieval of a NP's chirality $\alpha_{em}$ based on Eq. (5), is possible only by figuring out the helicity density at the position of the NP under test. For example, if we assume that the NP under test is located on the $z$ axis (i.e., the beam axis) on the surface of the proposed NA, where the helicity density is locally maximized, the helicity density $h$ at this position is [38]

$$h \approx h_{inc} + \frac{h_{inc}}{\pi a^3 \epsilon_0}\left[\frac{|\alpha_{ee}^{NA}|^2}{4\pi a^3 \epsilon_0} + \Re\{\alpha_{ee}^{NA}\} - \frac{a}{\lambda}\left(\frac{\pi^2+1}{\pi}\right)\Im\{\alpha_{ee}^{NA}\}\right], \tag{9}$$

with $h_{inc} = u_{inc}/\omega$ being the helicity density of the incident beam at the NA position. Therefore, by knowing the helicity of the incident light and the NA's electric polarizability, one obtains the NPs chirality $\alpha_{em}$ from Eq. (5).

It is important to note that the proposed NA is achiral since we do not want to include the NA's chirality contribution to the differential extinction. Indeed, it is hard to estimate and differentiate such a contribution from the differential extinction generated by the test chiral NP. In this discussion we have neglected the feedback mechanism due to the coupling between the chiral sample and NA since generally the polarizability of extremely subwavelength NPs is weak.

Next, we discuss the characterization of the NP chirality in terms of the magnetoelectric polarizability with respect to its electric polarizability. To this end, we consider the dissymmetry factor $g$ defined as [48]

$$g = \frac{\Delta P_{ext}}{\bar{P}_{ext}}, \tag{10}$$

where $\bar{P}_{ext} = (P_{ext}^+ + P_{ext}^-)/2$. For NPs with negligible magnetic properties, i.e., $\alpha_{mm} \approx 0$, Eq. (10) reads [21]

$$g = \frac{\Re\{\alpha_{em}\}}{\eta_0 \Im\{\alpha_{ee}\}} \frac{\omega h}{u_e}, \tag{11}$$

under conditions (4) and (6), where $u_e = u_e^+ = u_e^-$. In Eq. (11) the "field term" $\omega h / u_e$ is separated from the NP's property $\Re\{\alpha_{em}\}/(\eta_0 \Im\{\alpha_{ee}\})$. It is interesting to find the general property of structured lights that makes this field term equal to a universal constant value, and hence to eliminate the need of specific knowledge about energy and helicity densities in determining the dissymmetry factor $g$. This leads to empowering structured lights to exclusively reveal chirality of matter at nanoscale. Indeed, the introduced concept of helicity maximization summarized in conditions (7) does exactly that, i.e., it implies that $\omega |h|/u_e = 2$. This means that dissymmetry factor $g$ for such fields reduces to

$$g = \frac{2\Re\{\alpha_{em}\}}{\eta_0 \Im\{\alpha_{ee}\}}, \quad (12)$$

which is independent of the filed properties when optimally chiral structured light is used, and it provides exclusive information about a NP's chirality with respect to its electric property.

In conclusion, we have introduced the concept of helicity maximization applicable to any structured light, including optical beams and that arising from the near field of nanoantennas. We have discovered the upper bound of helicity density for a given energy density, obtained when conditions (7) are satisfied. Among the many choices of structured light that maximize helicity density, we have proposed the superposition of two special focused beams, i.e., the azimuthally polarized beam and the radially polarized beam, with a specific field ratio, and proper phase shift. Moreover, we have introduced the concept of nanoantennas that generates a near field with maximized helicity density (with an order of magnitude enhancement). This concept enables the elimination of field properties from dissymmetry factor $g$, and suggests the best possible use of nanoantennas for chirality detection and characterization. These findings have the capability to empower characterization of chirality at nanoscale by taking the advantage of field localization, structured light, and helicity maximization.

The authors acknowledge support by the W. M. Keck Foundation.